\documentclass[aps,prl,twocolumn,showpacs]{revtex4}
\usepackage{graphicx}
\usepackage{amsfonts,amssymb}
\usepackage{amsmath}
\DeclareGraphicsExtensions{.eps,.ps,.eps.gz,.ps.gz,.eps.Z,{}}

\input{psfig.sty}

\begin{document}

\title{Atom-molecule collisions in an optically trapped gas}

\author{N. Zahzam\footnote{Email adress: nassim.zahzam@lac.u-psud.fr}, T. Vogt, M. Mudrich\footnote{%
Present address: Physikalisches Institut, Universit\"{a}t Freiburg,
Germany}, D. Comparat and P. Pillet}
\affiliation{Laboratoire Aim\'e Cotton\footnote{Laboratoire Aim\'e Cotton is associated with Universit\'e
Paris-Sud. Web site: \texttt{www.lac.u-psud.fr}}, CNRS, Campus d'Orsay B\^at. 505, 91405 Orsay, France}

\begin{abstract}
Cold inelastic collisions between confined cesium (Cs) atoms and Cs$_2$ molecules are investigated inside a CO$_2$ laser dipole trap. Inelastic atom-molecule collisions can be observed and measured with a rate coefficient of $\sim 2.5 \times 10^{-11}\,$cm$^3$ s$^{-1}$, mainly independent of the molecular ro-vibrational state populated. Lifetimes of purely atomic and molecular samples are essentially limited by rest gas collisions. The pure molecular trap lifetime ranges 0,3-1 s, four times smaller than the atomic one, as is also observed in a pure magnetic trap. We give an estimation of the inelastic molecule-molecule collision rate to be $\sim 10^{-11}$ cm$^{3}$ s$^{-1}$.
\end{abstract}


\maketitle

Cold molecular gases have established a novel field of research in the past few years. After demonstrations of several methods to prepare cold samples of molecules, spectacular advances have been achieved, 
one of them being the formation of molecular condensates from atomic Fermi gases~\cite{Greiner,Jochim}. Increasing inter-disciplinary efforts are motivated by a wide range of new phenomena and applications~\cite{QuoVadis}. E.\,g., cold polar molecules are candidate systems for precision measurements~\cite{2002PhRvL..89b3003H} and for quantum information schemes~\cite{2002PhRvL..88f7901D}. 

Understanding and controlling the collisional properties of mixed atomic and molecular gases is crucial to achieve the regime of quantum degenerate molecular gases. Up to now, very few experimental data are available, and systematic studies of stored atoms, molecules or mixture of both, have just started~\cite{regal,Mukaiyama,Quemener}. 
As a recent highlight, the formation of ultracold Cs$_4$ molecules by Feshbach collisions of Cs$_2$ molecules has been observed~\cite{Grimm}.

Although a large variety of techniques for the production of cold molecular samples is currently being developed~\cite{QuoVadis}, sub-mK temperatures are only reached by photoassociation (PA) out of an initially ultracold atomic sample~\cite{Comparat}, and by molecule association through magnetic Feshbach resonances~\cite{Inouye}. The two techniques are complementary with respect to the nature of the formed molecules: While Feshbach-association leads to loosely bound molecules occupying the most weakly bound ro-vibrational levels, PA allows to populate deeper bound ro-vibrational states.

Cold molecular samples can be stored using magnetic trapping as demonstrated with Cs$_{2}$~\cite{VanHaecke} and KRb~\cite{Wang} in their triplet state. Trapping molecules in any state is possible in an optical trap as demonstrated by using a CO$_{2}$ laser (quasi-electrostatic trap, QUEST) with Cs$_{2}$ molecules present in a MOT~\cite{Takekoshi} or formed through a Feshbach resonance~\cite{Grimm} and with Rb$_{2}$ molecules formed via PA~\cite{Fioretti2}.

In this letter, we report the realization of a CO$_{2}$ laser based QUEST for mixed samples of Cs atoms and Cs$_2$ molecules. The Cs$_2$ molecules are formed in the electronic ground state or in the lowest triplet state via PA of cold
trapped atoms and are also efficiently trapped inside the QUEST. Lifetime measurements are performed for both atomic and molecular samples. We characterize the inelastic collisions between cold atoms, between cold atoms and cold molecules, and even between cold molecules inside the QUEST. Quantitative data analysis gives us access to the collision rate coefficients.

The basic experimental setup has been previously described (see e.\,g.~\cite{VanHaecke}). The main change here is the implementation of a CO$_{2}$ laser to realize the QUEST. The cold atoms are provided by a Cs vapor-loaded magneto-optical trap (MOT) with a residual gas pressure in the range of $10^{-7}-10^{-8}$ Pa. The shape of the atomic cloud is approximatively spherical, with a radius of $300$ $ \mu $m. The number of atoms in the MOT is $10^{7}$, leading to a peak density of $2\times 10^{10}$ cm$^{-3}$ by assuming a gaussian density distribution. A cw CO$_{2}$ laser (Synrad Firestar f100) is focused into the MOT zone with a waist of $\sim 80$ $\mu $m at an available power of $\sim 110$ W. The CO$_{2}$ laser is first roughly aligned with the cold atomic cloud through two ZnSe windows of the vacuum chamber by using a counter-propagating beam, provided by a diode laser beam tuned on the atomic transition $6s_{1/2},f=4\longrightarrow  6p_{3/2},f'=5$. For more accurate alignment, we use the decrease of MOT fluorescence  due to the light shift induced by the CO$_{2}$ laser. 

The temporal sequence for the loading of the QUEST from the MOT is the following. The CO$_{2}$ laser is permanently on. First we cool the atoms with an optical molasses phase of 15 ms by switching off the magnetic field gradient and by red-detuning the cooling laser from $2.5$ $\Gamma $ to $27$ $\Gamma$ relatively to the atomic transition $6s_{1/2},f=4 \longrightarrow  6p_{3/2},f'=5$ ; here $\Gamma \approx 2\pi \times 5.2$ MHz corresponds to the decay rate of the $6p_{3/2}$ state. The efficiency of the transfer of atoms from the MOT to the QUEST is typically around three percent. The detection of the atoms is performed by a resonant three-photon ionization process via the level $10s$, using a pulsed dye laser ($\lambda =706.7$ nm) pumped by the second harmonic of a Nd:YAG laser. To prepare the atoms in the hyperfine level $f=3$ (resp. $f=4$), we switch off the repumping beam 3\,ms before (resp. after) the end of the molasses phase. 

\begin{figure}[t]
\begin{center}
\includegraphics* [scale=0.8]{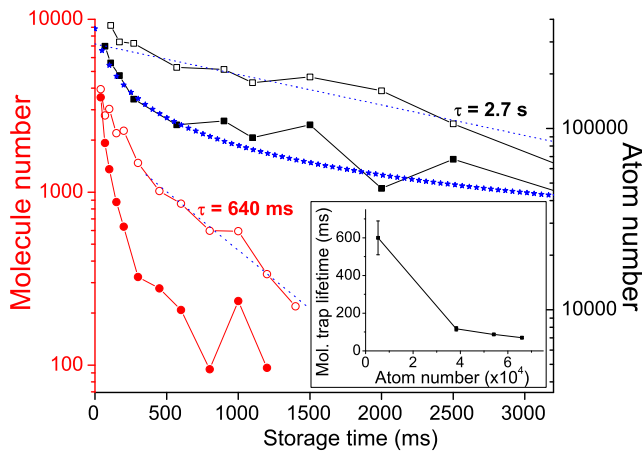} 
\caption{(Color online) Temporal evolution of the atomic and molecular population in the QUEST. The evolution of the number of trapped atoms prepared in the hyperfine states $f=3$ (open squares line) and $f=4$ (filled squares line) is plotted. The decay of atoms in $f=3$ is fitted by an exponentially decreasing function (dot line) with a time constant $\tau=2.7$\,s, whereas, for atoms in $f=4$, a fit function accounting for two-body collisions (filled stars) is necessary. The filled circles curve shows the evolution of the number of molecules in a mixed atomic and molecular trap. In the inset, we plot the decay constant associated with the molecular trap for different numbers of atoms left in the QUEST. We also represent the evolution of a pure molecular sample (open circles line), fitted by an exponential decay function (dotted line) with a time constant $\tau=640$\,ms. In these experiments, molecules are formed via PA of the state $0_g^-(6s+6p_{3/2})\;(v=6,J=2)$.}
\label{fig:at_mol}
\end{center}
\end{figure}

Fig. \ref{fig:at_mol} shows the temporal evolution of the number of atoms in the QUEST, $N_{\mathrm{Cs}}(t)$. For the hyperfine ground state level $f=3$, we observe an exponential drop with a lifetime of $2.7\pm 0.2$ s, limited by the background vacuum pressure. However for $f=4$ hyperfine level, inelastic two-body collisions lead to a clearly non-exponential decay at short times. Assuming a Gaussian distribution of the atoms and considering the binary collision parameter $G_{\mathrm{Cs}}=1.1\times 10^{-11}$\,cm$^{3}$s$^{-1}$~\cite{Mudrich}, we were able to determine an average density $\overline{n}_{\mathrm{Cs}}(0)\approx 3\times10^{11}$ cm$^{-3}$ after loading the trap. The corresponding initial number of atoms is $N_{\mathrm{Cs}}(0)\approx 3.5 \times 10^5$.

The cold trapped atoms are photoassociated in a chosen rovibrationnal level $(v,J)$, of state $\Omega _{u,g}^{+,-}$, converging towards the electronically excited limit $6s+6p$
\[
2\text{Cs}(6s,f) + h\nu_{PA} \longrightarrow \text{Cs}_{2}(\Omega
_{u,g}^{+,-}(6s+6p_{1/2,3/2});v,J)
\]
We have considered different states, $0_{g}^{-}$, $\ 1_{u}(6s+6p_{3/2})$
and $0_{u}^{+}(6s+6p_{1/2})$, leading, for $u$ symmetry (resp. $g$ symetry), to the formation of cold molecules in the ground state, $X^{1}\Sigma _{g}^{+}$ (resp. in the lowest triplet state, $a^{3}\Sigma _{u}^{+}$)~\cite{Dion}. The PA laser is provided by a Ti:Sapphire laser (Coherent 899\ ring laser) pumped by an Argon-ion laser.\ The laser beam is focused to a $\simeq 300$ $\mu $m spot with an available intensity of 300 W\,cm$^{-2}$. The maximum number of formed molecules in the QUEST is obtained by applying the PA laser during 30 ms. For longer times, the number of molecules decreases due to the excitation of the molecules by the PA laser. In order to detect the translationaly cold Cs$_{2}$ molecules, we photoionize them into Cs$_{2}^{+}$ ions (REMPI\ process), and selectively detected them with a pair of microchannel plates through a time-of-flight mass spectrometer. Molecular photoionization is provided by the same pulsed dye laser as the one used for atomic detection, but using a slightly different wavelength, around $712$ nm. We detect an initial number of molecules $N_{\mathrm{Cs_2}}(0)\approx$ 4000, corresponding to an average molecular density $\overline{n}_{\mathrm{Cs_2}}(0)\approx 10^{10}$ cm$^{-3}$.
\begin{figure}[t]
\begin{center}
\includegraphics* [scale=0.8]{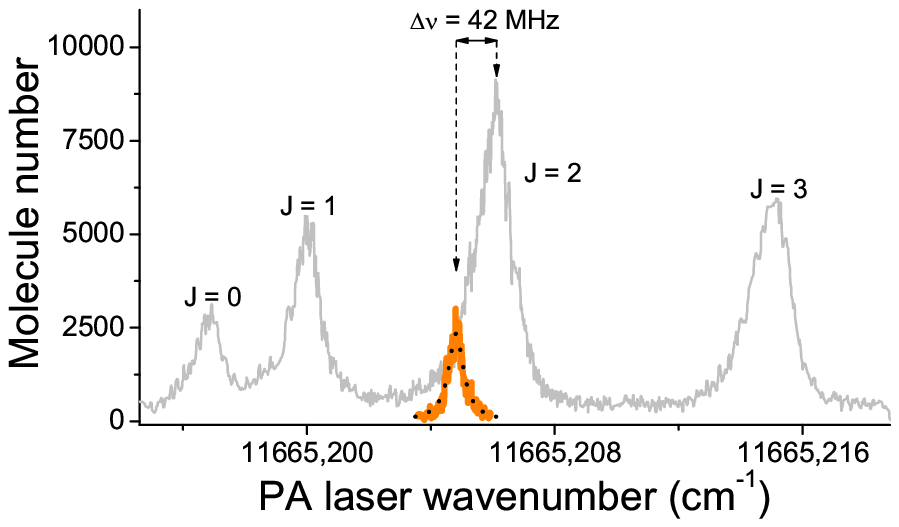} 
\caption{(Color online) PA spectrum of the $0_g^-(6s+6p_{3/2})$ $(v=6,J)$ lines in the MOT (grey line) and in the optical trap (dark line (orange online)). A fit of the latter is also represented (dotted line) (see text).
}
\label{fig:Scanv6}
\end{center}
\end{figure}

Fig.~\ref{fig:Scanv6} shows typical PA\ spectra obtained in a MOT and in a QUEST, corresponding to the $0_{g}^{-}(6s+6p_{3/2}) (v=6,J)$ excitation. For the MOT spectrum the rotational levels $J=0$ to $3$ are shown, and for the QUEST only $J=2$ level was recorded in this figure. We observe a 42\,MHz shift of the resonance between the two spectra because of the dynamical Stark effect due to the CO$_{2}$\ laser. The expected value of this shift is given by 
$
\Delta \nu =\left( \alpha _{6s}-\alpha _{6p_{3/2}}\right)P/(\pi h \varepsilon_{0} c w_0^{2}),
$
where $\alpha _{6s}$ and $\alpha_{6p_{3/2}}$ are the electrostatic polarizabilities of atoms in $6s$ and $6p_{3/2}$ states~\cite{1966PhRv..147...55M}, $P$ is the power of the laser beam and $w_0$ its waist. Considering an available power of 110\,W, we find $w_0=84\pm 4$\,$\mu$m, leading to an atomic potential depth of 0.9\,mK and a molecular one of 1.5\,mK, calculated with the Cs$_2$ static polarizability given in~\cite{tarnovsky}. Considering similar atomic QUEST experiments~\cite{Mudrich}, we assume an atomic trap temperature $T\approx 40_{-10}^{+10}\;\mu$K. We will also consider the molecular temperature to be roughly identical to the atomic one since the PA process does not induce any additional heating~\cite{Comparat}.\\
We want to mention here that the analysis of the PA line shape (see fit in Fig.~\ref{fig:Scanv6}) allows us to observe the magnetic sub levels structure of the $J=2$ state whose degeneracy is lifted by the CO$_2$ laser electric field~\cite{Kraft,Friedrich}. This leads to a PA line broadening of $\approx 5$ MHz in addition to the natural linewidth of $\approx$ 10 MHz. The full analysis of the PA lineshape is here in agreement with a temperature of $\approx 40\;\mu$K.  

Fig.~\ref{fig:at_mol} shows the temporal evolution of the number of atoms and molecules in the trap. The lower curve of this figure (filled circles) shows a lifetime of 110\,ms for the molecular cloud in presence of cold atoms, which is quite short compared to the 2.7\,s lifetime of the atomic cloud. This molecular trap lifetime does not depend on whether the atoms left in the trap are in $f=3$ or $f=4$ hyperfine state. Similar lifetime measurements are obtained after PA of atoms in any of those two hyperfine states. We interpret this lifetime as the result of inelastic collisions between cold atoms and cold molecules. By pushing the atoms outside the QUEST, using a resonant laser beam on the transition $6s_{1/2},f=4\longrightarrow 6p_{3/2},f^{\prime }=5$ (the repumping beam is on again during this phase), we observe an increase of this lifetime up to 640 ms. The inset of Fig.~\ref{fig:at_mol} shows the evolution of the molecular sample lifetime versus the number of atoms left in the QUEST. We can see that this lifetime is quite sensitive even to a small number of atoms in the trap. The dynamics of the molecular population for the mixed trap is given by the rate equation
\begin{align*}
\frac{dN_{\mathrm{Cs_2}}(t)}{dt}\simeq - \left( \Gamma_{\mathrm{Cs_2}}+K_{\mathrm{Cs_{2}Cs}}
\overline{n}_{\mathrm{Cs}}(t)\right) N_{\mathrm{Cs_2}}(t)&\\
-2G_{\mathrm{Cs_2}}\overline{n}_{\mathrm{Cs_2}}(t)N_{\mathrm{Cs_2}}(t)&,
\end{align*}
where $\Gamma_{\mathrm{Cs_2}}$ is the background loss rate, $K_{\mathrm{Cs_{2}Cs}}$ and $G_{\mathrm{Cs_2}}$ are the atom-molecule and molecule-molecule collision rate coefficient, respectively. Neglecting at early times $G_{\mathrm{Cs_2}}$, we extract from our data the parameter $K_{\mathrm{Cs_{2}Cs}}=2.5_{-0.5}^{+2}\times 10^{-11}$ cm$^{3}$ s$^{-1}$. This large incertainty on the value of $K_{\mathrm{Cs_{2}Cs}}$ comes mostly from the incertainty on the atomic temperature.
\begin{figure}[t]
\begin{center}
\includegraphics* [scale=0.8]{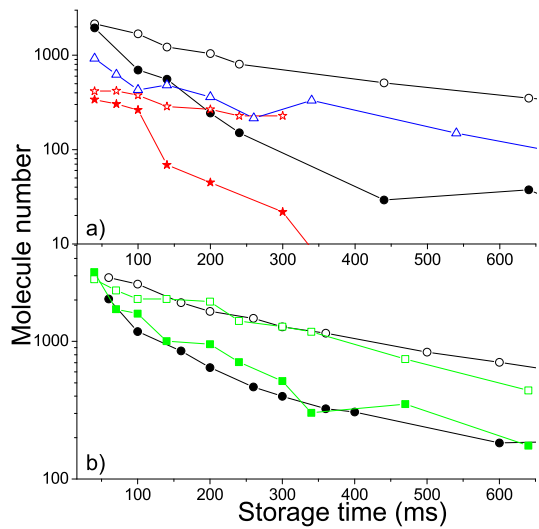} 
\caption{(Color online) a) Trap lifetime of a molecular sample with atoms (filled symbols) and without (open symbols), after PA of the $0_{g}^{-} (6s+6p_{3/2})$ (circles line), $1_{u}(6s+6p_{3/2})$ (triangles line) and $0_{u}^{+} (6s+6p_{1/2})$ (stars line) states. b) For different experimental conditions, the evolution of the number of trapped molecules with atoms (filled symbols) and without (open symbols) is represented after PA of the $v=6$ (circle symbols) and $v=103$ (square symbols) level of the $0_{g}^{-} ((6s+6p_{3/2})$ state.}
\label{fig:0g1u0u}
\end{center}
\end{figure}

As already mentioned, PA in the potentials $0_{g}^{-}$, $1_{u}(6s+6p_{3/2})$ and $0_{u}^{+}(6s+6p_{1/2})$ was also performed, as shown in Fig.~\ref{fig:0g1u0u}a. We have also considered a particular scheme for the formation of molecules (see Fig. \ref{fig:0g1u0u}b), where atoms are photoassociated into one level ($v=103$) of the outer well of the $0_{g}^{-}(6s+6p_{3/2})$ potential, coupled by tunneling effect with those of the inner one and predicted to form deeper bound molecules in the triplet ground state~\cite{Vatasescu}. Cold atom-molecule collision rate and molecule collision rate with background gas does not seem to depend on the photoassociated state.

All observed lifetimes of a pure molecular trap are about four times smaller than those observed for the atomic cloud ($f=3$) in the same conditions of vacuum background. To explain this difference, we have first tested the role of the CO$_{2}$ laser intensity on both trap lifetimes. We have considered two different waists (84 and 100\,$\mu $m) for the QUEST leading to two different laser intensities. We have not observed any difference for the atomic and molecular trap lifetime for both waists. It does not seem that the CO$_{2}$ laser induces significant transitions between ro-vibrational levels which are electric-dipole allowed due to the contribution of the second-order spin-orbit interaction \cite{Vanhaecke3}. It seems that the explanation for this factor 4 comes from the fact that the cross-section for collisions between hot atom and cold molecules is larger by the same factor compared to the hot atom-cold atom cross-section. This result is not so surprising if we consider that the formed and detected molecules are in a bunch of ro-vibrationnal levels close in energy to the dissociation limit $6s+6s$. These molecules have large maximum elongations leading to large impact parameters. To confirm this assertion, we have measured the atomic and molecular trap lifetimes in the case of a pure magnetic confinement with a setup similar to the one of \cite{Vanhaecke3}. Here the atomic and molecular densities are two orders of magnitude lower, and the inelastic collisions between cold species can be ignored. We have seen that both lifetimes are limited by the background gas collisions and are mostly identical to those observed in the QUEST, with the same factor 4 between atomic and molecular case.
\begin{figure}[t]
\begin{center}
\includegraphics* [scale=0.8]{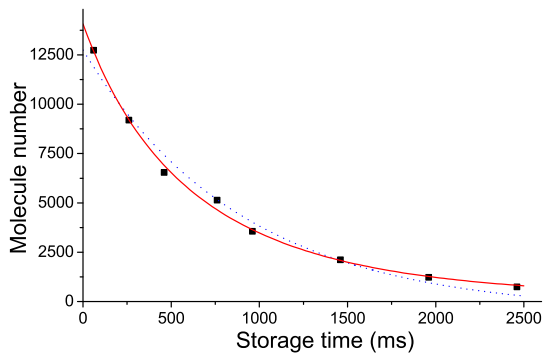} 
\caption{(Color online) Trap lifetime of a pure molecular sample. The evolution of the number of molecules, formed via PA of the $0_g^-(6s+6p_{3/2})$ $(v=6,J=2)$ state, is plotted (filled squares) and fitted with a function accounting for two-body collisions (solid line). A fit with an exponential decay function is also represented (dotted line). The fit yields a lifetime of 900\,ms and a two-body rate coefficient of $G_{\mathrm{Cs_2}}= 1.0(5)\times 10^{-11}$ cm$^{3}$ s$^{-1}$.}
\label{fig:molmol}
\end{center}
\end{figure}

The role of molecule-molecule collisions is not dominant in our experiment compared to the one of background gas collisions. Fig.~\ref{fig:molmol} gives the evolution of a pure molecular sample with optimized initial number of molecules and background gas pressure. A simple exponential decay function does not fit very well the data. Nevertheless, with a two-body fit function, the data are better ajusted and we estimate $G_{\mathrm{Cs_2}}= 1.0(5)\times 10^{-11}$\,cm$^{3}$ s$^{-1}$, in agreement with results given in reference~\cite{Grimm}.

To conclude, we have reported the realization and the characterization of an atomic and molecular QUEST which constitutes a quite exciting tool to trap homonuclear molecules. Depending on the initial experimental conditions, we obtained cold molecule clouds with a number of molecules up to 12000 ($\overline{n}_{\mathrm{Cs_2}}(0)\approx 3\times10^{10}$ cm$^{-3}$) and trap lifetimes up to 1\,s. A systematic collisional study inside the QUEST allowed us to determine the atom-molecule and molecule-molecule collision rates. Both rates are close to the unitarity limit ($\approx 1.5 \times 10^{-11}$ cm$^{3}$ s$^{-1}$ for atom-molecule collisions and $\approx 8 \times 10^{-12}$ cm$^{3}$ s$^{-1}$ for molecule-molecule collisions) and are not compatible with the formation of a stable Bose-Einstein condensate of molecules formed initially from bosonic atoms. The theoretical reference~\cite{1998PhRvL..80.3224B} predicts an atom-molecule collisional rate quite sensitive to the vibrational level of the formed molecule. In our experiment, we have not seen any evidence of such dependence, but further experiments with selected ro-vibrationnal levels are necessary to definitely conclude. Concerning the lifetime of a pure molecular trap, molecule collisions with background gas is the limiting process; the collision cross-section beeing about a factor 4 higher than the one for the atoms. To go further, it can be interesting to increase the trap lifetime by decreasing the vacuum background pressure, which should permit us to measure the lifetime of the metastable molecular triplet state $a^{3}\Sigma _{u}^{+}$, to study more precisely molecule-molecule collisions and to observe molecular transitions induced by the CO$_2$ laser.

Up to now, the molecules are preferentially formed in high vibrational levels. An interesting challenge is to prepare the molecules in low vibrational levels of the singlet state which may exhibit different behavior against collisional processes. Two-photon PA~\cite{2000Sci...287.1016W,vanhaecke2} toward highly excited states can offer good configuration. Another challenge is to prepare the molecules in a well defined ro-vibrationnal level using Raman PA~\cite{2000Sci...287.1016W,vanhaecke2}, chirped pulse PA~\cite{koch} or microwave PA. At term a dense and cold molecular sample could be prepared in the lowest rovibrational level ($v=0,J=0$) of the ground state where Bose-Einstein condensation of such a gas would not be impeded by inelastic collisions. 

\acknowledgments M.M.\ was supported as postdoct by the COMOL network (contract HPRN-CT-2002-00309). The authors thank N. Vanhaecke for preliminary design of the laser trap, and A. Fioretti, O. Dulieu, A. Crubellier and F. Masnou-Seeuws for stimulating discussions.

\textit{Note added:} We recently became aware that similar work has been carried out simultaneously in the group of M. Weidem\"{u}ller at the Physikalisches Institut in Freiburg, Germany.

\bibliographystyle{apsrev}

\end{document}